\documentclass[a4paper,12pt]{article} 
\usepackage{epsfig}
\usepackage{graphicx}

\newcommand{\beq}{\begin{eqnarray}}
\newcommand{\eeq}{\end{eqnarray}}

\begin{document}
\pagestyle{plain}
%\pagestyle{empty}

%
%%%%%%%%%%%%%%%%%%%%%%%%%%%%%%%%%%%%%%%%%%%%%%%%%%%%%%%%%%%%%%%%%%%%%%
% Titre et auteur
%
\title{
{\Large \bf 
Studies of beam offset due to beam-beam interactions at a warm linear collider
}
}
 
\author{
{\bf 
Nicolas Delerue\footnote{\tt n.delerue1@physics.ox.ac.uk}\ \footnote{Now at the
 John Adams Institute for Accelerator Science at the University of
 Oxford},
Toshiaki Tauchi\footnote{\tt toshiaki.tauchi@kek.jp}
 and 
Kaoru Yokoya\footnote{\tt kaoru.yokoya@kek.jp}
} \\ 
\em High Energy Accelerator Research Organization (KEK),\\
\em 1-1 Oho, Tsukuba Science City, 305-0801 Ibaraki-ken, Japan
}

%\date{
%July 2004
%}

%\twocolumn[
%\vspace*{-2cm}
\maketitle
%\thispagestyle{empty}

%\vspace*{-1cm}
%%%%%%%%%%%%%%%%%%%%%%%%%%%%%%%%%%%%%%%%%%%%%%%%%%%%%%%%%%%%%%%%%%%%%%
% Resume
%
\begin{center}

{\bf 
At a warm linear collider the short time interval at which bunches
will pass near each other in the interaction region may lead to
significant alteration of the bunches positions.\\ In this paper we quantify
the intensity of this effect and show that it can be addressed by a fast intra-pulse feedback system.
} \\ 
\vspace*{.5cm}
{\it To be submitted to Physical Review Special Topics, Accelerators
  and Beams }
\end{center}
%]

\vspace*{.5cm}

%
%---------------------------------------------------------------------
\section{Beam-beam interaction and beam blow up at a warm linear collider}
%---------------------------------------------------------------------
%

\begin{figure}[hbtp]
\begin{tabular}{p{7cm}p{6cm}}
\centering\epsfig{file=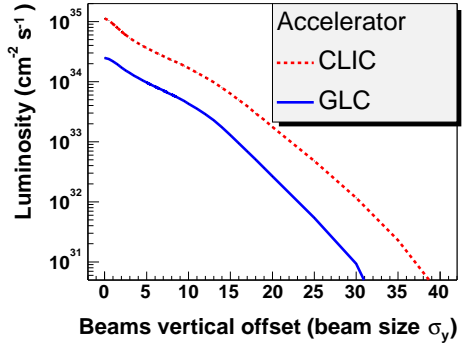,width=7.cm}&
\vspace*{-5.5cm}
\begin{tabular}{p{6cm}}
\caption{Total luminosity delivered as a function of the
  vertical offset of the beams at the interaction point. The
  horizontal unit, $\sigma_y$, is the vertical size of the beam (a few
  nanometers). The parameters used for this simulation are those of
  the GLC and of CLIC as given in
  table~\ref{tab:parametersValue} page~\pageref{tab:parametersValue}. Perfect crab-crossing (or head-on
  collisions have been assumed).}
\label{fig:lumiremain} 
\end{tabular}
\end{tabular}
\end{figure}

In a linear collider, near the interaction point (IP) after the final magnet the two beams are
not any more shielded from each other by the beam pipe. Thus if the
outgoing beam has been deflected vertically at the interaction point it will
induce a vertical deflection of the incoming beam, leading to a
loss of luminosity (see figure~\ref{fig:lumiremain}) and an increasing
displacement of
the beam along the train.

%
%---------------FIGURE
%

\begin{figure}[hbtp]
\begin{tabular}{p{7cm}p{6cm}}
\centering{\epsfig{file=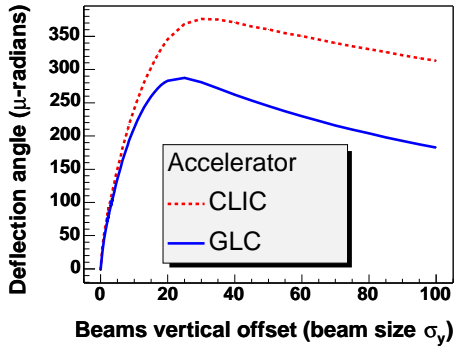,width=7.cm}} &
\vspace*{-4.5cm}
\begin{tabular}{p{6cm}}
\caption{
Beam deflection angle ($\theta_y$) as a function the bunch offset at the
interaction point ($\Delta_{k}$). The parameters used for this simulation are those of
  the GLC and of CLIC as given in table~\ref{tab:parametersValue}.}
\label{fig:formFactor} 
\end{tabular}
\end{tabular}
\end{figure}

 Let the vertical offset of the $k$-th electron (positron) bunch
at the IP in units of the rms.~beam size $\sigma_y$ be $\Delta_k^{(-)}$
($\Delta_k^{(+)}$) and define the relative offset by
$\Delta_k=\Delta_k^{(-)}-\Delta_k^{(+)}$ (we assume that the two beam
have roughly the same size).
Let the offsets without beam-beam interaction be $\Delta_{k,0}$.
Then, the offsets with interaction are obtained successively
by\cite{Yokoya:1991qz} 
\beq
   \Delta_k = C \sum_{l=\max(1,(k-N))}^{k-1} F(\Delta_l) +
      \Delta_{k,0}  \label{eq:blowup} \\
      \qquad  C \equiv \left(\frac{\sigma_x}{\sigma_z \phi}\right)^2
      D_x D_y \label{eq:C} 
\eeq
where $D_{x(y)}$ is the horizontal (vertical) disruption parameter,
$\sigma_{x(z)}$ the horizontal (longitudinal) bunch size, $\phi$ the
crossing angle, $N$ the number of bunches that a given bunch sees on its
journey
from the last quad to the IP. The form factor $F(\Delta)$ is defined by
\begin{equation}
     F(\Delta) = \frac{\gamma(\sigma_x+\sigma_y)}{N_pr_e} \theta_y(\Delta)
\end{equation}
where $r_e$ is the classical electron radius, $N_p$ the number of
particles in a bunch, $\gamma$ the particle energy in units of rest
mass and $\theta_y(\Delta)$ the beam deflection angle when the beam
offset
is $\Delta$, as shown on figure~\ref{fig:formFactor}. ($F(\Delta)\approx \Delta$, when $D_y\ll 1$ and
$\vert\Delta\vert\ll 1$.)

%According to calculations made by 
% Kaoru Yokoya and
%Pisin Chen\cite{Yokoya:1991qz} 
%the offset ($\Delta_k$) of the $k^{th}$ bunch is given by the
% following relation:
%\beq
%\Delta_k & = & C \sum_{l=max(1,(k-N))}^{k-1} F(\Delta_l) + \Delta_{k,0} \\
%C & = & \frac{D_x D_y}{c^2_{\phi}} \label{eq:C}
%\eeq
%where 
%$\Delta_{k,0}$ is the initial offset of the incoming $k^{th}$ bunch, 
%$\Delta_l$ is the offset of the outgoing bunch $l$,
%$F(\Delta_l)$ is a form factor shown on figure~\ref{fig:formFactor} (simulated with CAIN~\cite{Chen:1995jt})  relating the deflection of a given bunch
% with its offset,
%$D_x$ and $D_y$ are the vertical and horizontal disruption parameters
% of the beam, 
%$c_{\phi}$ is the crossing angle between the two beams 
%and $N$ is
% the number of bunches that a given bunch ``sees'' on its journey
% from the last quad to the interaction point.

This number $N$ can be calculated using the following formula:
\beq
N & = & \frac{2 \times \mbox{distance between IP and  last quad}}{\mbox{Bunch spacing} \times c}  \label{eq:N}
\eeq
(here $c$ is the velocity of the beam taken as the velocity of the
light in the calculations below).

%
%---------------FIGURE
%

An interesting point to note in this formula is that the offset is
independent of the location at which bunch $k$ and $l$ cross each
other. This happens because two different effects compensate each
other. On the one hand the further away from the IP the crossing
happens, the bigger the distance between bunches $k$ and $l$ is
and thus the smaller the deflection angle of the bunch $k$ will be. But one the
other hand, the distance traveled by the bunch after receiving this
kick will be longer, thus making the offset at the IP bigger.

The simulations presented in this paper have been done using
CAIN~\cite{Chen:1995jt} with two sets of parameters, one close
to the proposed parameters of the GLC (Global Linear Collider) and
the other closer to the CLIC specifications.
The parameters' values used for these studies are adopted from the ITRC
report~\cite{unknown:2003na} and are summarized in
table~\ref{tab:parametersValue}.

%
%---------------FIGURE
%
\begin{table}[hbtp]
\begin{center}
\begin{tabular}{c|c|c|}
Set & GLC/NLC & CLIC \\
\hline
Energy (GeV) &	243 & 202 \\
\hline
$\sigma_x$ (nm) &	243 & 202 \\
$\sigma_y$ (nm) & 3 & 1.2 \\
$\sigma_{y'}$ ($\mu$rad) & 27 & 24 \\
$\sigma_z$ ($\mu$m) & 110 & 35 \\
$D_x$ &	0.16 & 0.04 \\
$D_y$ & 13.1 & 6.4 \\
$C$ [eq~\ref{eq:C}] & 0.209 & 0.0213 \\ 
$\phi$ (crossing angle) (mrad) & 7 (20) & 20 \\
\hline
Bunch spacing (ns) & 1.4 & 0.67 \\
$L^*$ (distance between IP and last quad) (m) & 3.5 & 4.3 \\
N (bunches) [eq~\ref{eq:N}] & 16 & 42 \\
\hline
\hline
\end{tabular}
\caption{Beam parameter values (at 500 GeV) used from the blow up simulations
  adopted from the ITRC report~\cite{unknown:2003na}. }
\label{tab:parametersValue} 
\end{center}
\end{table}
%\end{figure}
%

%
%---------------------------------------------------------------------
\section{Effect of the crossing angle and the other beam parameters on
  the beam blow up}
%---------------------------------------------------------------------
%

\begin{figure}[p]
\begin{center}
\vspace*{-4cm}
\hspace*{-2cm}\epsfig{file=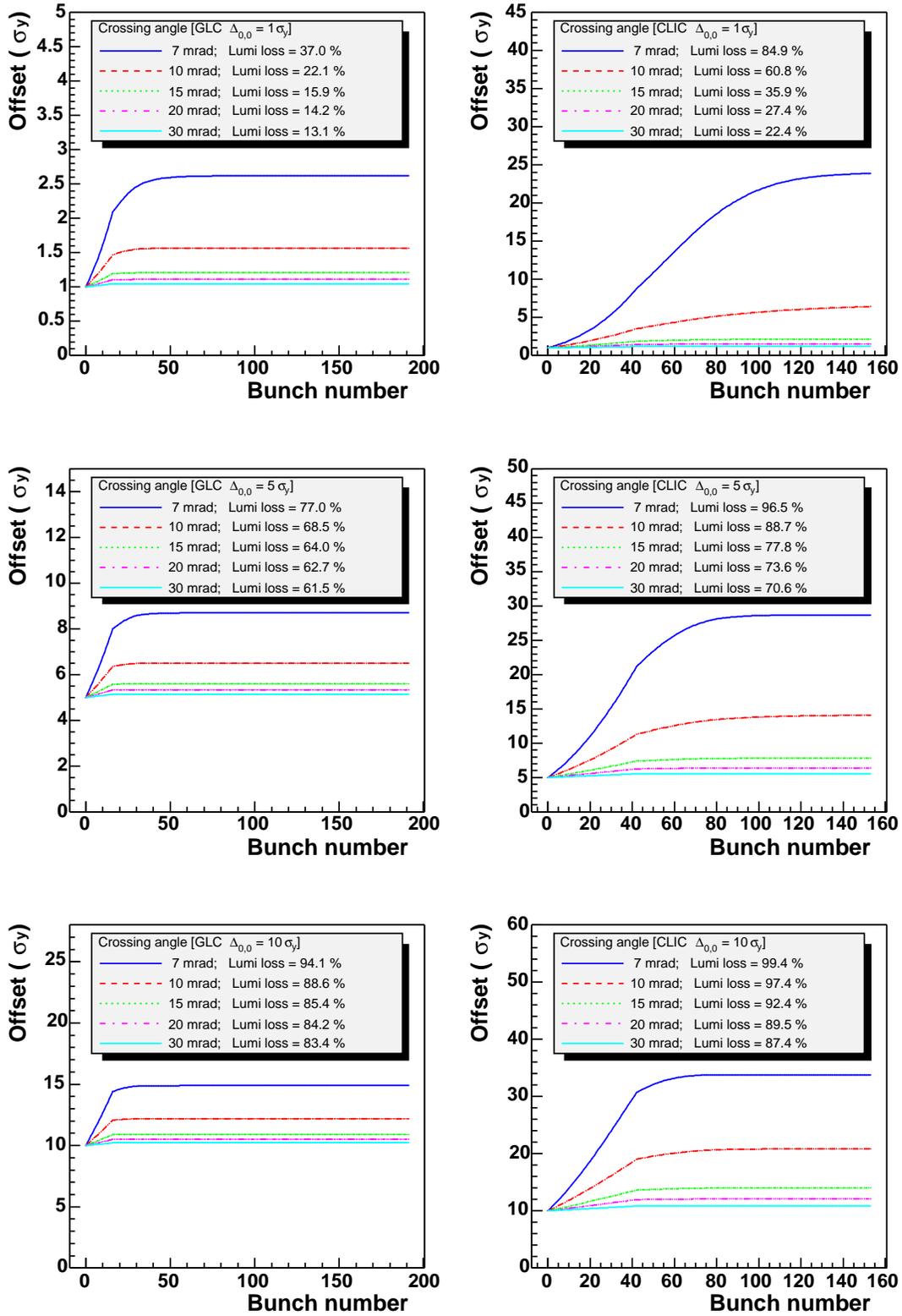,width=17cm}
\vspace*{-1cm}
\caption{Beam blow up as a function of the bunch number for an initial
  beam offset of $1 \sigma_y$ (upper plots), $5 \sigma_y$ (middle plots) or $10
  \sigma_y$ (lower plots) for different values of the crossing
  angle. The left column correspond to simulations with the GLC (NLC)
  parameters and the right column correspond to simulations with the
  CLIC parameters.}
\label{fig:blowupEffect} 
\end{center}
\end{figure}

The two parameters that have the biggest influence on the beam blow up
are the crossing angle and the number (N) of outgoing bunches seen by an
incoming bunch. The figure \ref{fig:blowupEffect} shows how the blow
up (simulated as described by equation~\ref{eq:blowup}) varies when the crossing angle varies from the smallest proposed
value (7~mrad) to a much less challenging value (30 mrad) and the
table \ref{fig:blowupEffectValues} indicates the vertical offset of the
last bunch of the train (It is assumed that $\Delta_{k,0}$ is the same for all
bunches of a train). 

%This assumption is justified by the fact that time between
%two trains is many order of magnitudes ($\sim 5$) longer than the
%length of the train, thus the variation of $\Delta_{k,0}$ within a
%train is many orders of magnitude smaller than its variation from
%train to train.

\begin{figure}[p]
\vspace*{-4cm}
\begin{center}
\hspace*{-2cm}\epsfig{file=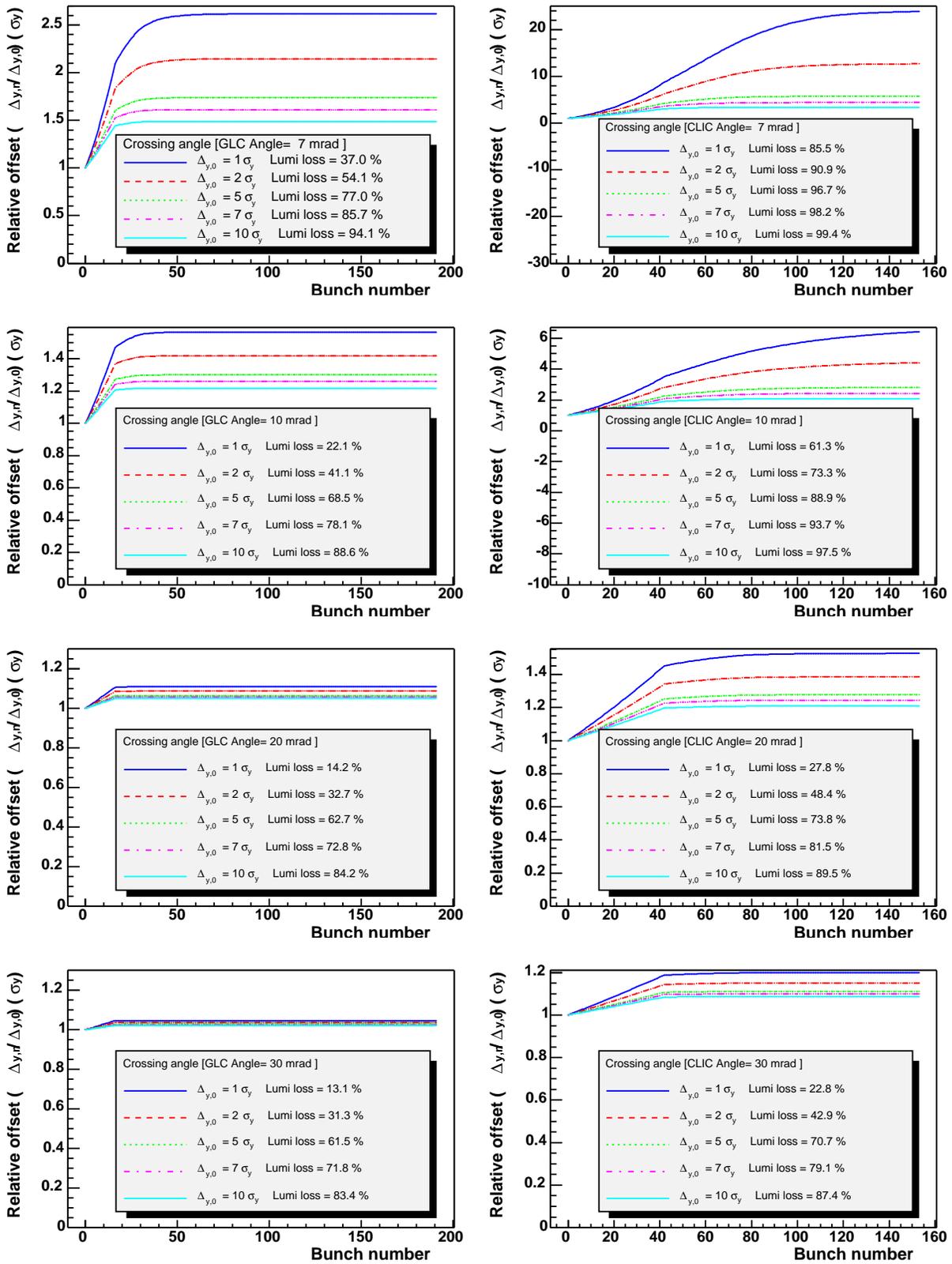,width=18cm}
\vspace*{-2cm}
\caption{Normalized beam blow up as a function of the bunch number for
  a crossing angle of (from top to bottom) 7~mrad, 10~mrad,
  20~mrad and 30~mrad for different
  values of the initial beam offset. 
The normalization is done by dividing the actual
  offset of a given bunch by the initial beam offset.}
\label{fig:blowupEffectNormalized} 
\end{center}
\end{figure}

As one can see on this figure, for an initial offset of $1 \sigma_y$,
even with a crossing angle of 7~mrad the maximum beam offset at the
GLC due to the beam blow up does not exceed 3 $\sigma_y$. For a crossing
angle of 10 mrad or more the beam offset remains below 1.6 $\sigma_y$.
At CLIC the shorter bunch spacing increases the blow up effect. It
can reach 23.9~$\sigma_y$ for a crossing angle of 7~mrad and 5.3~$\sigma_y$ for a crossing angle of 10 mrad. For wider crossing angle,
the blow up remains below 2 $\sigma_y$. If the initial offset is
bigger (5~$\sigma_y$ or 10~$\sigma_y$) then the final offset increases
but the increase, which is related to the form factor shown on
figure~\ref{fig:formFactor}, is less than linear and the normalized
offset ($\frac{\mbox{bunch offset}}{\mbox{initial offset}}$) is
  smaller, as shown on figure~\ref{fig:blowupEffectNormalized}).

%
%---------------FIGURE
%
%

%
%---------------FIGURE
%
%

%
%---------------FIGURE
%
\begin{table}[hbtp]
\begin{center}
\begin{tabular}{c||c|c||c|c||c|c||}
Crossing angle       & \multicolumn{2}{c||}{$1 \sigma_y$ offset} & \multicolumn{2}{c||}{$5 \sigma_y$ offset} & \multicolumn{2}{c||}{$10 \sigma_y$ offset} \\
($\phi$) (mrad)  & GLC/NLC & CLIC & GLC/NLC & CLIC & GLC/NLC & CLIC \\
\hline
 7 & 2.6 & 23.9 & 8.7 & 28.7 & 14.9 & 33.8 \\
10 & 1.6 &  6.4 & 6.5 & 14.1 & 12.2 & 20.8 \\
15 & 1.2 &  2.2 & 5.6 &  7.8 & 10.9 & 14.0 \\ 
20 & 1.1 &  1.5 & 5.3 &  6.3 & 10.5 & 12.1 \\
30 & 1.0 &  1.2 & 5.1 &  5.6 & 10.2 & 10.9 \\
\hline
\hline
\end{tabular}
\caption{Vertical offset (expressed in beam size $\sigma_y$) of the last  of the train for various
  crossing angles (the other parameters have the values mentioned in
  table~\ref{tab:parametersValue}).
}
\label{fig:blowupEffectValues}
\end{center}
\end{table}

To minimize the blowup effect it is better to locate the last quad
 closer
      to the IP but this is not desirable for the detector performance.
      Instead, it may be possible to keep
the incoming and outgoing beam in separate pipes until they are very
close from the IP, thus shielding them for each other's influence and
reducing~N, the number of outgoing bunches seen by an incoming bunch (and
thus the blow up).

\begin{figure}[hbtp]
\hspace*{-1.5cm}\epsfig{file=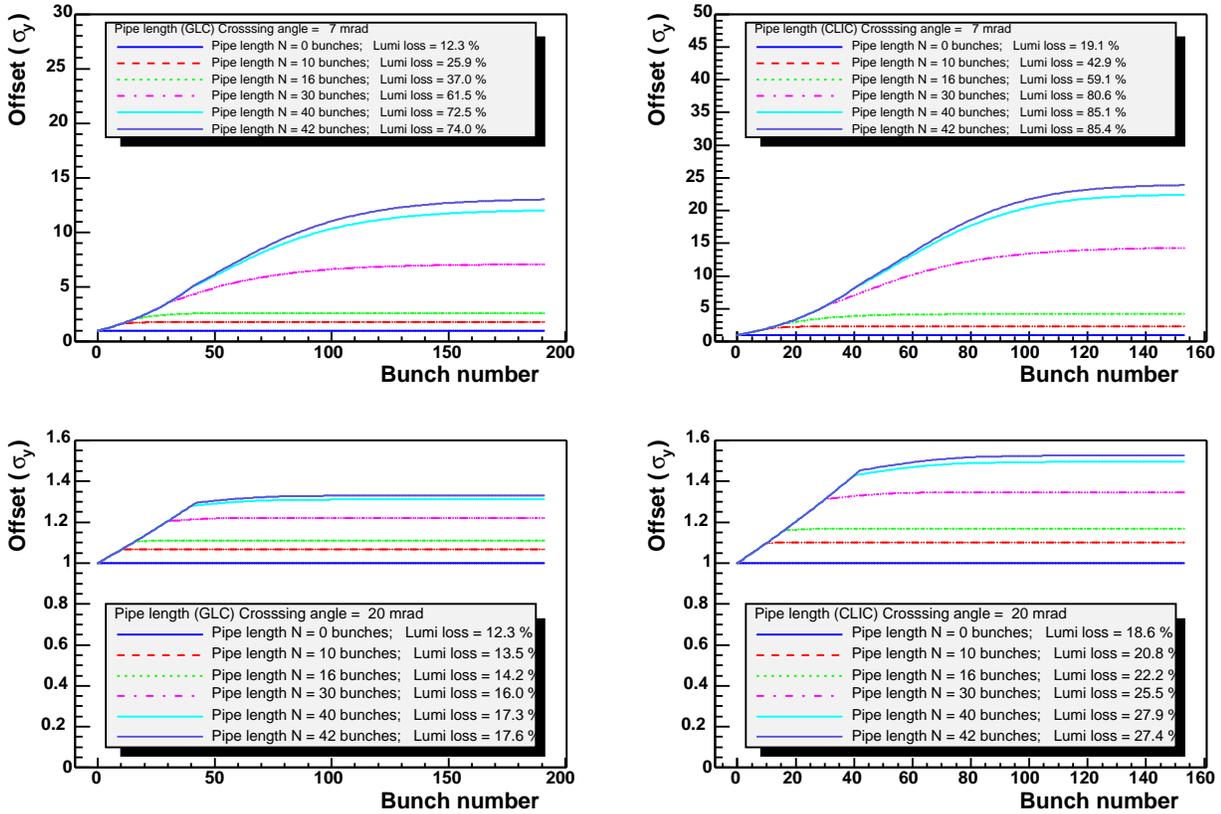,width=17cm}
\caption{Beam blow up as a function of the number of outgoing bunches
  seen by an incoming bunch for an initial
  beam offset of $1 \sigma_y$ for a crossing angle of 7~mrad (upper plots)
  and 20 mrad (lower plots).}
\label{fig:blowupPipe} 
\end{figure}

On figure \ref{fig:blowupPipe} one can see the influence of N (number
of outgoing bunches seen by an incoming bunch) on the blow up.
The offset of
the last bunch of the train for various values of N is given in table~\ref{fig:blowupPipeValues}. As shown
previously, at very small crossing angle (7~mrad), the beam blow up is very
important and thus the unshielded length has a strong influence on the
total blow up. With such crossing angle reducing the unshielded length by 20 cm,
from 4.2~m to 4.0~m can reduce the vertical offset of the last bunch
of the train by more than $1.0 \sigma_y$ for both the GLC and CLIC.  
At higher crossing angle (20 mrad), the beam blow up is
much lower and thus the unshielded length has a smaller influence on
the blow up.

%
%---------------FIGURE
%
%

%
%---------------FIGURE
%
\begin{table}[hbtp]
\begin{center}
\begin{tabular}{c||c|c||c|c||c|c||}
Number of     & \multicolumn{2}{c||}{IR length $L^{*}$ (m)} & \multicolumn{2}{c||}{7 mrad} & \multicolumn{2}{c||}{20 mrad} \\
bunches seen  & GLC/NLC & CLIC & GLC/NLC & CLIC & GLC/NLC & CLIC \\
\hline
 0 & $0 \pm 0.2$ & $0 \pm 0.1$ &  1   & 1    & 1   & 1   \\
10 & $2.1 \pm 0.2$ & $1.0 \pm 0.1$ &  1.8 & 2.3  & 1.1 & 1.1 \\
16 & $3.4 \pm 0.2$ & $1.6 \pm 0.1$ &  2.6 & 4.2  & 1.1 & 1.2 \\
30 & $6.3 \pm 0.2$ & $3.0 \pm 0.1$ &  7.1 & 14.3 & 1.2 & 1.3 \\
40 & $8.4 \pm 0.2$ & $4.0 \pm 0.1$ & 12.0 & 22.4 & 1.3 & 1.5 \\
42 & $8.8 \pm 0.2$ & $4.2 \pm 0.1$ & 13.0 & 23.9 & 1.3 & 1.5  \\
\hline
\hline
\end{tabular}
\caption{Vertical offset (expressed in beam size, $\sigma_y$) of the beams at the
  end of the train for different values of N (number of outgoing bunches seen by an
  incoming bunch) for two different crossing angle values (the other parameters have the values mentioned in
  table~\ref{tab:parametersValue}). N is proportional to the length
  during which the two beam are not shielded from each other in the
  interaction region (IR).
}
\label{fig:blowupPipeValues}
\end{center}
\end{table}
%

%
%---------------------------------------------------------------------
\section{Blow up and fast feedback system}
%---------------------------------------------------------------------
%

Ground motion and other sources of vibrations may induce random changes in
the beam offset from train. To reduce the luminosity lost due to this
offset  fast intra-pulse
feedback systems have been proposed for the GLC (and the
NLC)\cite{Delerue:2003xg,Schulte:2000ax,Burrows:2001qj} to correct the beam offset by
measuring the offset of the deflected outgoing pulses with a beam
position monitor (BPM) and correcting the
incoming pulses with a kicker. After initial correction a delay loop
acts to prevent the system from forgetting the correction already applied.

As the beam blow up also results in a beam offset, the
fast intrapulse feedback  systems can also deal with it. 

Mathematically the effect ($d_{k}$) of the feedback system described
in~\cite{Delerue:2003xg} on bunch $k$ can be described as follow:
\beq
\mbox{if } (k < b) \mbox{   then   } \delta_k & = & 0 \mbox{ and } d_k = 0 \\ 
\mbox{if } (k\ge b) \mbox{   then   } \delta_k & = & g * F(\Delta_{k-b}) \\
\mbox{ and } d_k & = & \delta_k + d_{k-b} \label{eq:feedback}
\eeq

Where $\Delta_{k,0}$ is the initial offset of the $k$-th bunch (as
defined above), $F(\Delta_{k})$ is the relative angle with which bunch
$k$ was deflected (F is shown on figure~\ref{fig:formFactor}), $b$ is
the latency of the system (that is the distance separating the BPM of the feedback
system from the kicker plus the electronic latency, expressed in number of bunches), $g$ is the gain of the system (typically 0.6
for the GLC in normal conditions) and $\delta_k$ is the correction specific to
bunch $k$ to which the
correction $d_{k-b}$ (memorized by the delay loop) is added to give
$d_{k}$, the
total correction to be applied.

\begin{figure}[hbtp]
\begin{center}
\epsfig{file=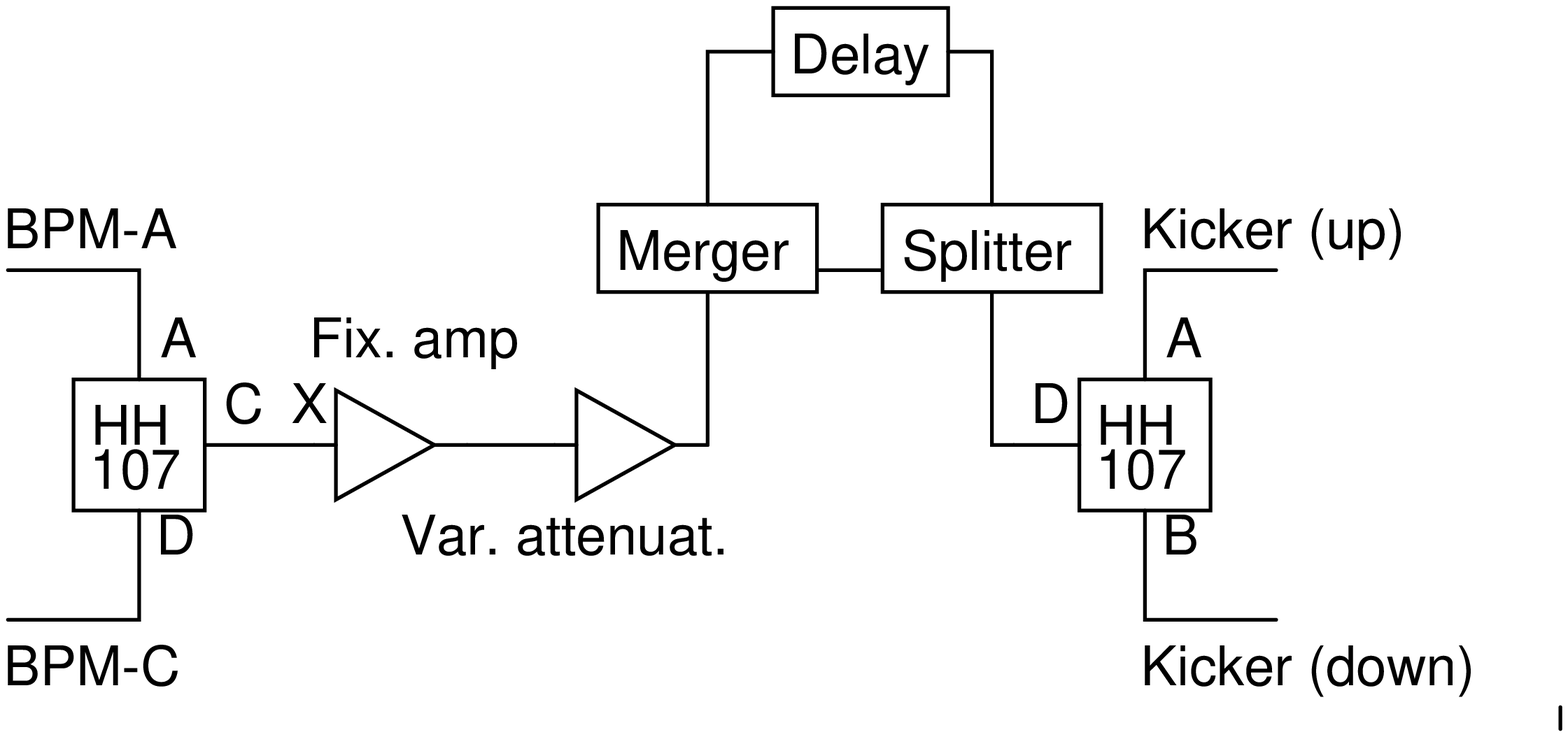,width=12.cm}
\caption{Circuit of the FEATHER\cite{Delerue:2003xg} fast intra-pulse feedback system.}
\label{fig:circuitDelayed} 
\end{center}
\end{figure}

This correction $d_k$ is directly subtracted from $\Delta_{k},0$
before computing the effect of the beam beam blow up as shown on
equation~\ref{eq:blowup}. The electronic circuit of the system
described in~\cite{Delerue:2003xg} is shown on figure~\ref{fig:circuitDelayed}.

%
%---------------FIGURE
%

\begin{figure}[hbtp]
\begin{center}
\vspace*{-2cm}
\epsfig{file=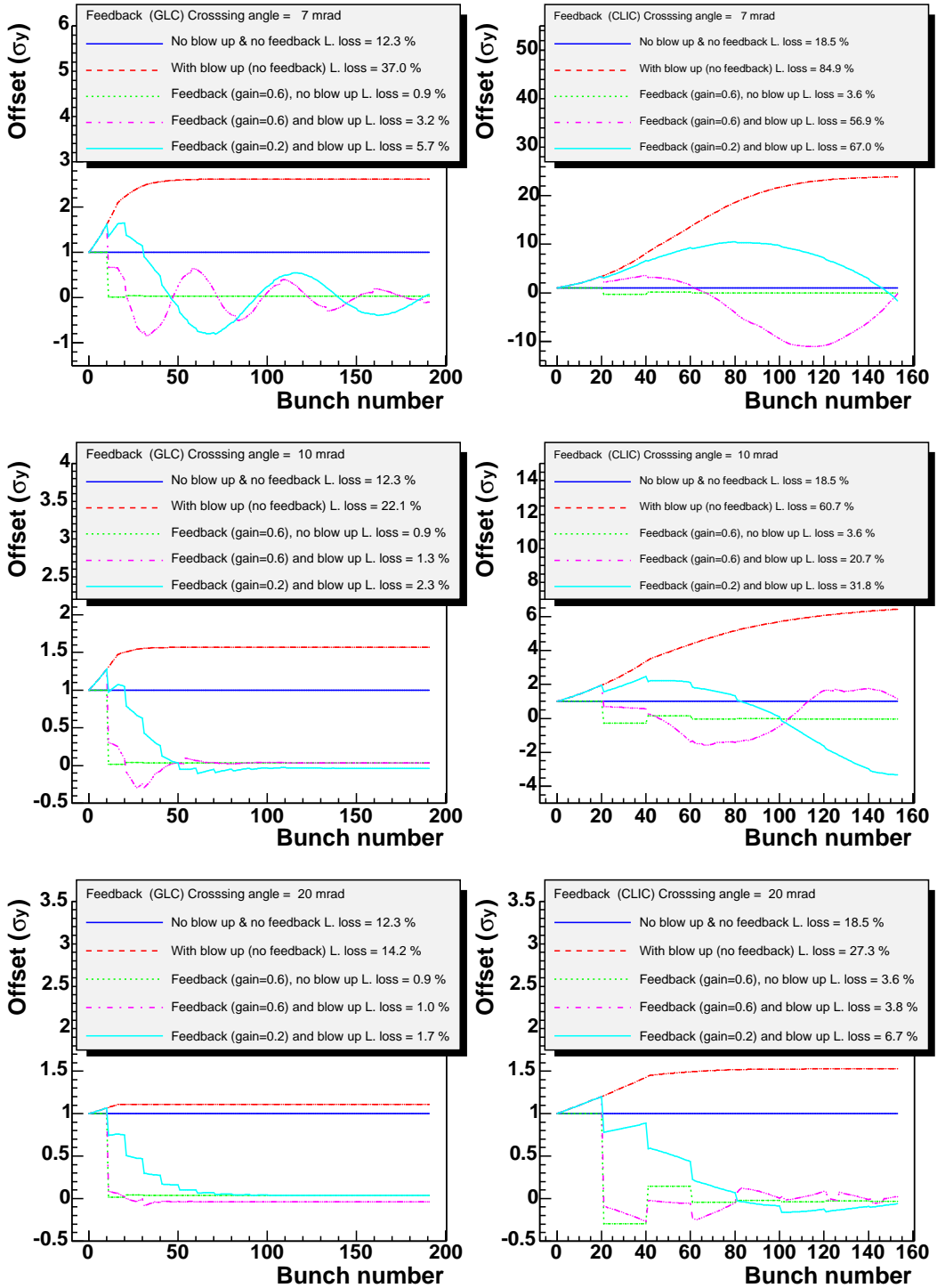,width=15.5cm}
\caption{Effect of the fast intra-pulse feedback system on the beam
  blow up for different crossing angles for the GLC (left column) and
  CLIC (right column) for an initial
  beam offset of $1 \sigma_y$. The feedback system is assumed to be
  located near the last quad.}
\label{fig:blowupFeedback} 
\end{center}
\end{figure}

The effect of the beam blow up on the performance of these fast
intra-pulse feedback systems is shown on
figure~\ref{fig:blowupFeedback}. As one can see when the crossing
angle is wide and thus blow up is
not too intense (10 mrad and more for the GLC, 20 mrad and more for
CLIC), the fast feedback system can correct the beam blow up whereas for
smaller crossing angles the blow up drives the feedback system into
oscillations between over-correction and under-correction.

%
%---------------FIGURE
%

\begin{figure}[htbp]
\begin{center}
\hspace*{-.8cm}\epsfig{file=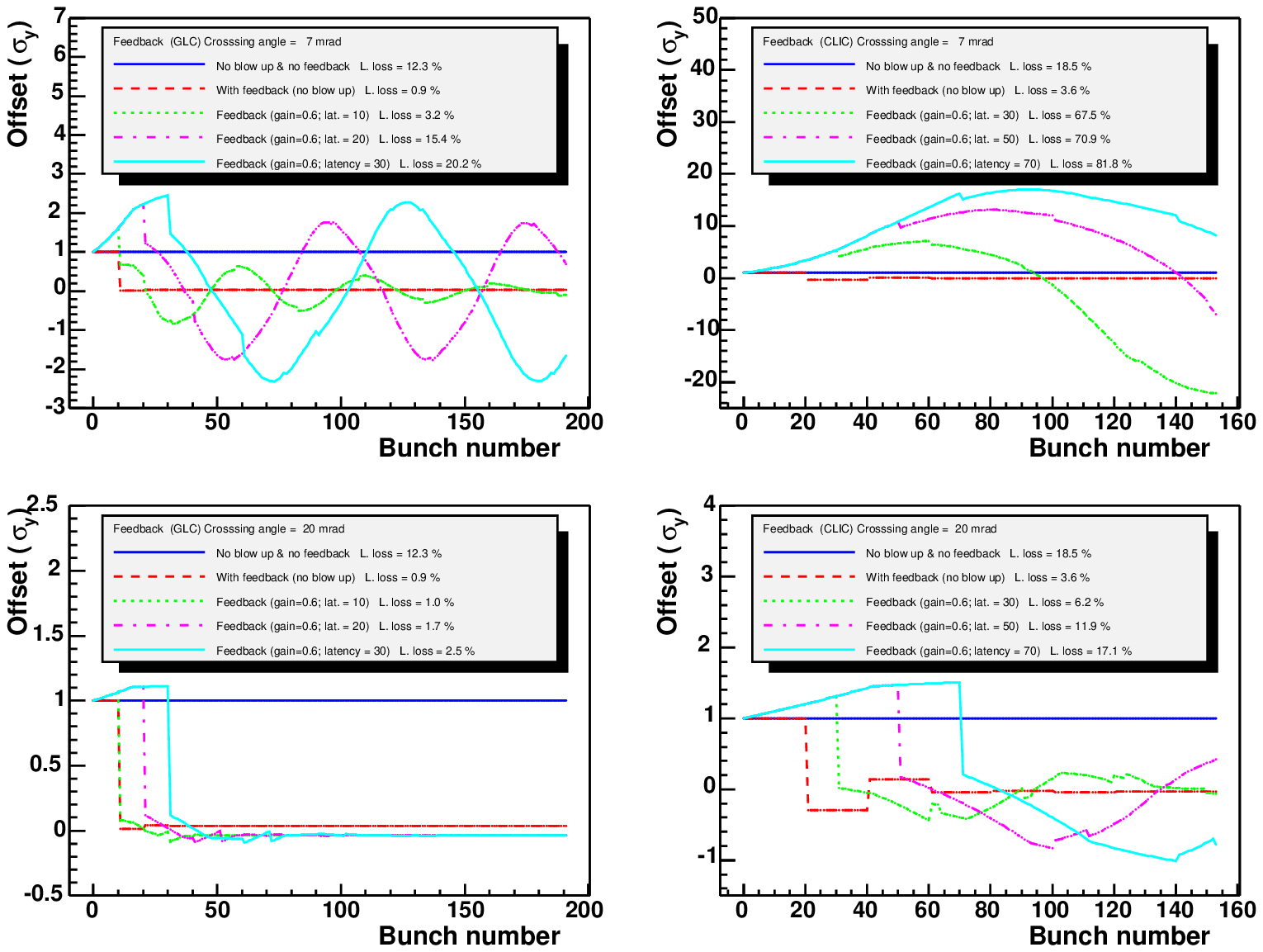,width=15.cm}
\caption{Effect of the fast intra-pulse feedback system on the beam
  blow up for different total latency (i.e. position) of the feedback system for the GLC and CLIC for an initial
  beam offset of $1 \sigma_y$. A feedback system latency of 10 (30) for the GLC
  corresponds to a feedback system located $\sim$2m ($\sim$6m) away from the IP. Latencies of 30
  (70) for CLIC corresponds to a feedback system located $\sim$2.8m ($\sim$5.8m) away from the IP.}
\label{fig:blowupFeedbackPosition} 
\end{center}
\end{figure}

These oscillations come from the delay between the time at which a
correction is applied and the time at which the BPM measures the
effects of this correction. Thus after correcting for a given effect
the system still measure ``uncorrected'' bunches. This delay is
induced by the time of flight from the kicker to the BPM and by the
latency of the electronics used.
The figure~\ref{fig:blowupFeedbackPosition}  shows that these
oscillations appear regardless of the position of the fast feedback
system (or the latency induced by the electronics), but their intensity
increases when the system is located further away from the IP.

%
%---------------FIGURE
%

The delay loop of the fast feedback system (see
figure~\ref{fig:circuitDelayed}) addresses some of the problems
created by the system's latency but it slows the capacity of the
system to adapt to changing conditions such as those created by the
beam blow up.

\begin{figure}[hbtp]
\begin{center}
\epsfig{file=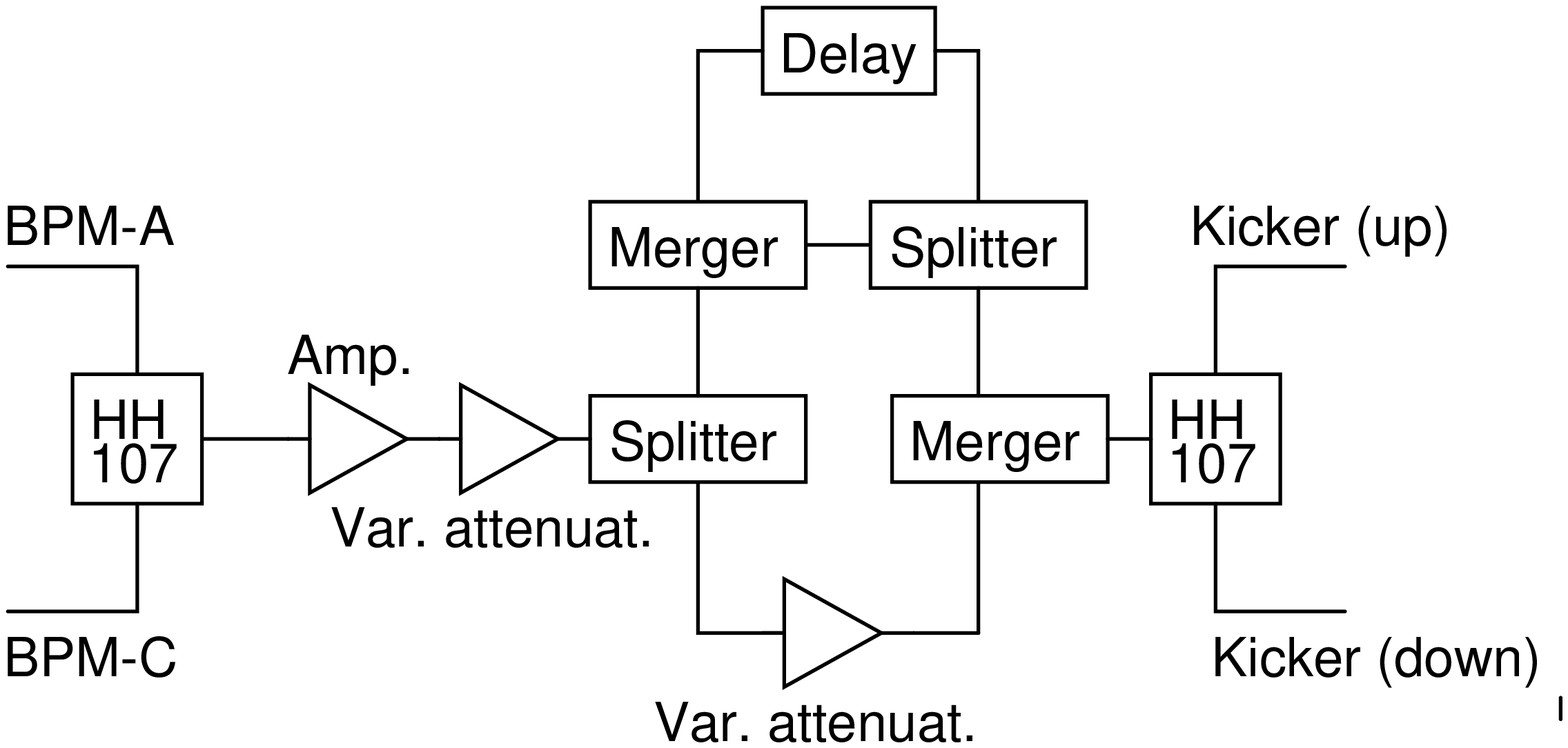,width=11.5cm}
\caption{Modified fast intra-pulse feedback system
  to avoid the oscillations created by the beam blow up. Compared
  to the FEATHER\cite{Delerue:2003xg} circuit one extra branch has
  been added that directly takes the beam position signal and bypasses
  the delay loop.}
\label{fig:circuitBlowup} 
\end{center}
\end{figure}

Thus to avoid the oscillations in the fast feedback system, one needs
to add a second component to the correction predicted by the feedback
system. The intensity of this second component must be directly
proportional to the measured bunch position and should not be included
in the delay loop. A modified feedback system including this second
component is shown on figure~\ref{fig:circuitBlowup}. 

%
%---------------FIGURE
%

Mathematically this new circuit requires the addition of a new term
$\epsilon_k$ to equation~\ref{eq:feedback} to compute the correction $c_k$:

\beq
\mbox{if } (k < b) \mbox{   then   } \delta_k & = & d_k = \epsilon_k =
c_k = 0 \\ 
\mbox{if } (k\ge b) \mbox{   then   } \delta_k & = & g * F(\Delta_{k-b}) \\
\mbox{ and } \epsilon_k & = & g_b * F(\Delta_{k-b}) \\
\mbox{ and } d_k & = & \delta_k + d_{k-b} \\
\mbox{ and } c_k & = & d_k + \epsilon_{k} \label{eq:feedbackBlowup}
\eeq

Where $c_k$ is the correction to be applied and $g_b$ is a
proportionality coefficient (the gain of the feedback branch). The only difference
between $\delta_k$ and $\epsilon_k$ is that the later is not included
in the recursive term $d_k$.

\begin{figure}[hbtp]
\begin{center}
%\vspace*{-2.cm}
\hspace*{-.8cm}\epsfig{file=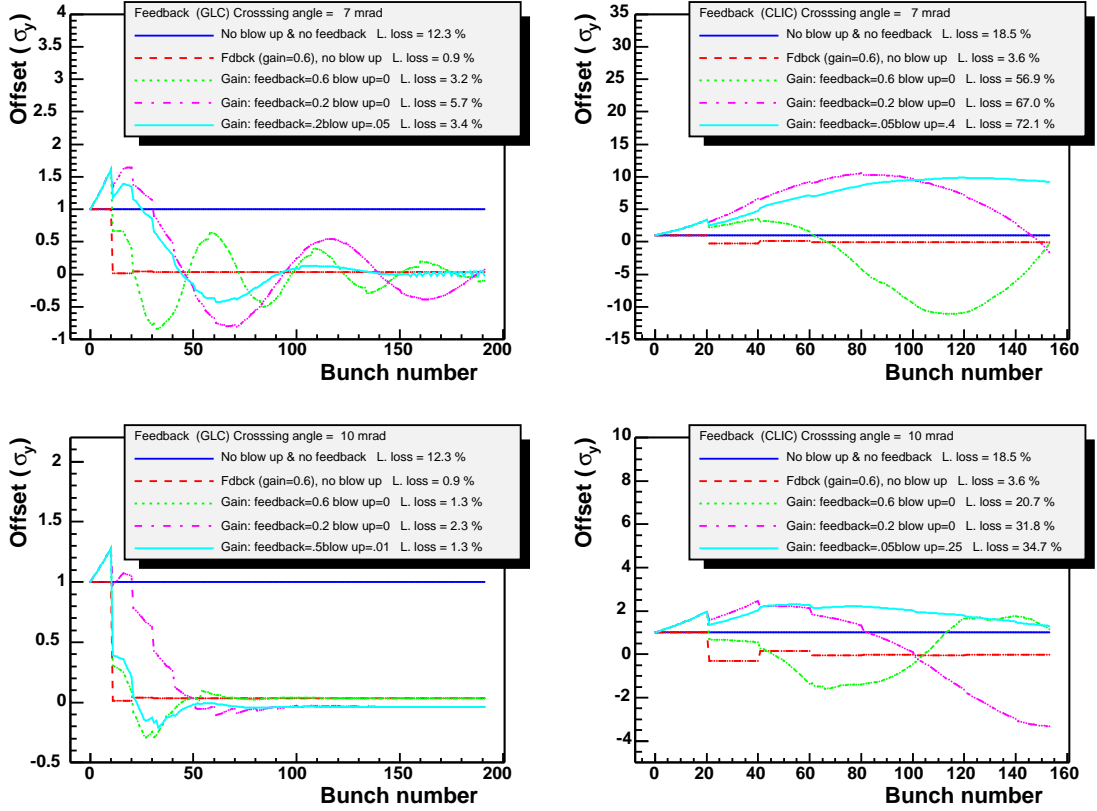,width=15.cm}
\caption{Effect of the modified fast intra-pulse feedback system on the beam
  blow up for different crossing angles for the GLC (left column) and
  CLIC (right column)  for an initial
  beam offset of $1 \sigma_y$. The feedback system is assumed to be
  located near the last quad. The fast feedback system used here has
  been modified as shown on figure~\ref{fig:circuitBlowup}. 
}
\label{fig:blowupFeedbackCorrected} 
\end{center}
\end{figure}

The performances  of this modified circuit are shown on
figure~\ref{fig:blowupFeedbackCorrected}. The gains
  used for these numerical simulations are $g=0.2$  and
  $g_b=1$  for the GLC with a crossing angle of 7 mrad, $g=0.05$
   and $g_b=18$ for CLIC at the same crossing
  angle. With a crossing angle of 10~mrad these values become $g=0.5$/$g_b=0.16$
  (GLC) and $g=0.05$/$g_b=11.25$ (CLIC). These values have been obtained by
  tuning the system to minimize the luminosity loss. The ratio between
  these two values reflects the contribution of the beam blow up to
  the total beam offset. 
As one can see this modification cancels or reduces the luminosity loss
due to the blow up.

%
%---------------FIGURE
%
\clearpage
%
%---------------------------------------------------------------------
\section{Blow up correction based only on the first bunch measurement}
%---------------------------------------------------------------------
%

\begin{figure}[hbtp]
\begin{center}
\epsfig{file=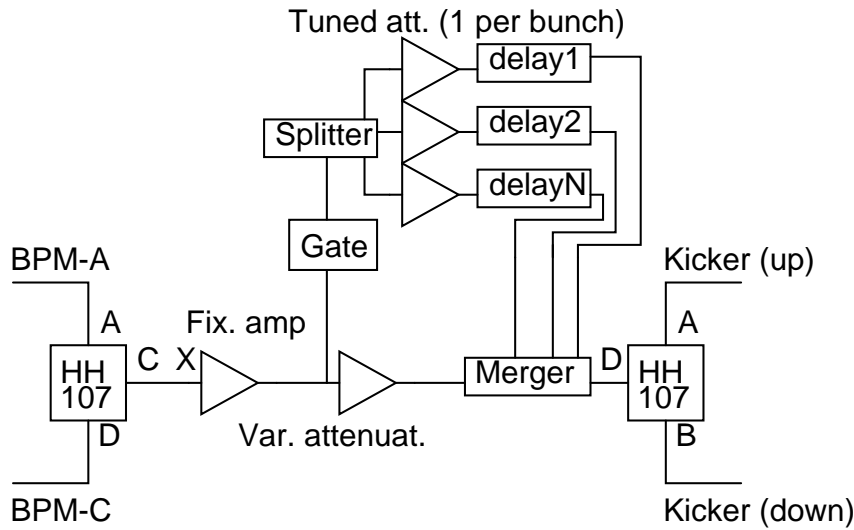,width=12.cm}
\caption{This circuit uses mainly the information coming
  from the first bunch to correct the whole train. To correct the
  residual components of the beam offset a simple feedback loop is
  also included.}
\label{fig:circuitAtOnce} 
\end{center}
\end{figure}

\begin{figure}[hbtp]
\begin{center}
\hspace*{-1.5cm}\epsfig{file=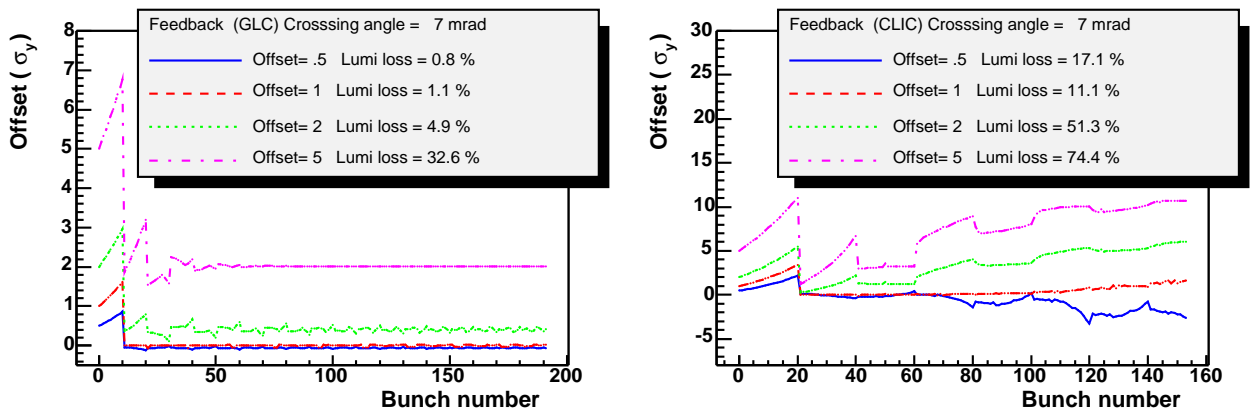,width=17.cm}
\caption{Correction of the beam offset using the circuit shown on
  figure~\ref{fig:circuitAtOnce}. The gain used for the simple feedback loop
  is 0.5 for the GLC and 0.7 for CLIC. The gains used in the other part
  of the circuit have been tuned so that the beam offset becomes null
  if the initial offset was 1.
}
\label{fig:blowupAtOnce} 
\end{center}
\end{figure}

As the offset of each bunch of the train is only affected by events
(ground motion, transverse long-range wakefield, compensation error of
beam-loading,...) that are known once the first bunch of the train reaches the IP, the correction
to be applied to each bunch can be predicted once the offset of
the first bunch is known.
This property could be used to design a
system that would compute the correction to be applied to
each bunch based mainly on the measurement of the offset of the first
bunch of the train.
To cope with residual ground motion, a simple feedback system (without
delay loop) must be
added to this system. As the correction to be applied as a function of
the bunch offset is not linear such system would have to be tuned for
a given offset at which it would perform the best. The simple feedback
loop would then perform the second order adjustments to remove the
residual beam offset.
By using switches
it would be possible to switch between different sets of gains tuned for
different initial offsets. The figure~\ref{fig:circuitAtOnce} shows an
example of circuit (without switch) that could be used to
implement such system.
The figure~\ref{fig:blowupAtOnce} shows the performances of such circuit.

%
%---------------FIGURE
%
%

%
%---------------FIGURE
%

 As one can see the performances of such system are rather attractive
 but the number of attenuators and
wires required would be proportional to the number of bunches times
the number of switches needed (as each of the bunches would require its
 own circuit). This huge number of wires needed might
be a problem as it would increase the amount of dead material in the detector.

It is important to stress
that in the two models presented in the previous section and the
one in this section only analog electronics have been used but by the date at
which the linear collider will be built very fast digital electronics will
probably be available allowing a better correction of the end of the
train.

%
%---------------------------------------------------------------------
\section{Conclusion}
%---------------------------------------------------------------------
%

The beam beam interactions at a warm linear collider such as the GLC
or CLIC will create a blow up of the beam, especially at low crossing
angle. If the crossing angle is wide enough then the blow up will be
corrected by the fast intra-pulse feedback system. For smaller crossing
angle the blow up will interfere with the feedback system but minor
modifications of the feedback system will remove these interferences
and correct the beam blow up.

%
%---------------------------------------------------------------------
\section{Acknowledgements}
%---------------------------------------------------------------------
%

One of the authors (ND) would like to thank JSPS 
for funding his stay in Japan under contract P02794.

\bibliographystyle{myunsrt}
\bibliography{biblio}

\end{document}